\documentclass[11pt]{article}

\pdfoutput=1
\usepackage[pdftex]{color}
\usepackage{amssymb}
\usepackage{amsthm}
\usepackage{amsmath}
\usepackage{latexsym}
\usepackage{epsfig}
\usepackage{amscd}
\usepackage{graphicx}
\usepackage[pdftex, colorlinks=true, citecolor=green]{hyperref}
\usepackage{lscape}
\usepackage{setspace}
\usepackage{cancel}
\usepackage{enumerate}

\renewcommand{\proof}[1]{\noindent \normalfont{\textbf{Proof.} \  \  #1 \hfill $\Box$}\\}

\newtheorem{thm}{Theorem}[section]
\newtheorem{prop}[thm]{Proposition}

\newcommand{\ds}{\displaystyle}

\begin{document}

\title{Input Statistics and Hebbian Crosstalk Effects}

\author{Anca R\v{a}dulescu$^1$}

\maketitle

$^1$Department of Mathematics, 395 UCB, University of Colorado, Boulder, 80309-0395\\

\hspace{3mm} phone: (303)492-6617; fax: (303)492-7707; email: \emph{radulesc@colorado.edu}\\

\vspace{5mm}
\begin{abstract}

\noindent As an extension of prior work, we study inspecific Hebbian learning using the classical Oja model. We use a combination of analytical tools and numerical simulations to investigate how the effects of inspecificity (or synaptic ``cross-talk'') depend on the input statistics. We investigated a variety of patterns that appear in dimensions higher than 2 (and classified them based on covariance type and input bias). The effects of inspecificity on the learning outcome were found to depend very strongly on the nature of the input, and in some cases were very dramatic, making unlikely the existence of a generic neural algorithm to correct learning inaccuracy due to cross-talk. We discuss the possibility that sophisticated learning, such as presumably occurs in the neocortex, is enabled as much by special \emph{proofreading} machinery for enhancing specificity, as by special algorithms.

\end{abstract}

\vspace{5mm}

\vspace{1cm}
\section{Introduction}

In this paper, we revisit some fundamental questions in computational neuroscience, related to unsupervised learning in cortical networks. We use a simple model of learning (previously studied by the author) to study how learning occurs when the model incorporates transmission inspecificity (``synaptic errors''). We focus in particular on a few crucial questions: To what extent and under which circumstances can synaptic inspecificity facilitate or prevent learning? Are certain input distributions more easily learned than others, or more affected by inspecificity? Can a small level of cross-talk induce significant changes (bifurcations) in the system's asymptotic dynamics?

The paper is organized as follows. In the introduction, we present the model (which we will call throughout the paper the ``inspecific Oja model'') and we overview the basics of its dynamic behavior (Section \ref{review}). In Section \ref{3dim} we start by investigating numerically how a 3-dimensional Oja inspecific network processes different classes of input distributions, preserving some of the dynamical aspects found in the 2-dimensional phase plane~\cite{radulescu2010}, but also introducing new features, specific to higher dimensions. In Section \ref{ndim}, we study analytically, in an $n$-dimensional example, the behavior observed numerically in the previous section. The Section \ref{disc} interprets the numerical and analytical results in the biological context of a learning cortical network.

\subsection{The inspecific Oja model}
\label{review}

Oja~\cite{oja1982simplified} showed that a simple neuronal model could perform unsupervised learning based on Hebbian synaptic weight updates  incorporating an implicit ``multiplicative'' weight normalization, to prevent unlimited weight growth~\cite{malsburg1973self}.

Our focus is on studying a single-output network, learning an input distribution according to Oja's rule~\cite{oja1982simplified}. More precisely, the output neuron receives, through a set of $n$ input neurons, $n$ signals ${\bf x} = (x_1,...,x_n)^T$ drawn from an input distribution ${\cal{P}}({\bf x}), \; {\bf x} \in \mathbb{R}^{n}$, transmitted via synaptic connections of strengths $\mbox{\boldmath $\omega$} = (\omega_1,...,\omega_n)^T$. The resulting scalar output $y$ is generated as the weighted sum of the inputs $y={\bf x}^T\mbox{\boldmath $\omega$}$. The synaptic weights $\omega_i$ are modified by implementing first a Hebb-like strengthening proportional to the product of $x_i$ and $y$ , followed by an approximate ``normalization'' step, maintaining the Euclidean norm of the weight vector close to one. The input covariance matrix ${\bf C}={\bf x}^T{\bf x}$ can be used as an appropriate long-term characterization of the inputs, to study the expected long-term convergence of the weight vector (i.e., learning), by approximating it with the asymptotic behavior of ${\bf w}(t)= \langle \mbox{\boldmath $\omega$}(t+1) \vert \mbox{\boldmath $\omega$}(t) \rangle$ in the equation~\cite{radulescu2009hebbian}:

$$ \frac{d{\bf w}}{dt} = \gamma \left[ {\bf C w} - \left( {\bf w}^{T}{\bf C w} \right) {\bf w} \right]$$

Oja~\cite{oja1982simplified} showed that this simple model acts, when applicable, as a principal component analyzer for the input distribution. Finding principal components could be very useful in the brain for data compression and transmission, since for Gaussian data such representations have statistically optimal properties, and often neural signals are approximately Gaussian.\\

\noindent Recent data suggest~\cite{harvey2007locally,bi2002spatiotemporal,bonhoeffer1989synaptic} that weight updates may be affected by each other, for example due to unavoidable residual second messenger diffusion between closely spaced synapses. In our recent work we examined how such ``crosstalk'' would affect the Oja model~\cite{radulescu2009hebbian}. We formalized learning inspecificity via an error matrix ${\bf E} \in {\cal{M}}_{n}(\mathbb{R})$ that has positive entries, is symmetric and equal to the identity matrix ${\bf }I \in {\cal{M}}_{n}(\mathbb{R})$ in case the error is zero. Consistently with our previous studies in both two~\cite{radulescu2010} and higher dimensions~\cite{radulescu2009hebbian}, we consider the error matrix of the form

\begin{equation}
{\bf E}=\left[\begin{array}{cccc}
q&\epsilon&\cdots &\epsilon\\
\epsilon&q&\cdots &\epsilon\\
\vdots & &\ddots &\vdots \\
\epsilon&\epsilon&\cdots &q\\
\end{array}\right]
\label{error}
\end{equation}

\noindent where $\displaystyle{0 < \epsilon < \frac{1}{n}}$ is the ``transmission error'' and $\displaystyle{\frac{1}{n} < q < 1}$ is the ``transmission quality,'' satisfying $q+(n-1)\epsilon=1$. The inspecific learning equations become:

\begin{equation}
\frac{d{\bf w}}{dt} = \gamma[{\bf ECw}-({\bf w}^{T}{\bf Cw}){\bf w}]
\label{mothersys}
\end{equation}

We have noted previously that an equilibrium for this system is any vector ${\bf w}=(w_{1}...w_{n})^{T}$ such that ${\bf ECw}=({\bf w}^{T}{\bf Cw}){\bf w}$, i.e., an eigenvector of ${\bf EC}$ (with corresponding eigenvalue $\lambda_{\bf w}$), normalized, with respect to the norm $\lVert \cdot \rVert_{\bf C} = \sqrt{\langle \cdot,\cdot \rangle_{\bf C}}$ (defined as $\langle {\bf v},{\bf u} \rangle_{\bf C}={\bf v}^T{\bf C}{\bf u}$, for all ${\bf u}, {\bf v} \in \mathbb{R}^n$), so that $\lVert {\bf w} \rVert_{\bf C}=\lambda_{\bf w}$. Generically, ${\bf EC}$ has a strictly positive, unique maximal eigenvalue, and the corresponding eigendirection is orthogonal in $\langle \langle \cdot,\cdot \rangle \rangle_{\bf C}$ to all other eigenvectors of ${\bf EC}$.

We have also shown that the eigenvalues of the Jacobian matrix at an equilibrium ${\bf w}$ are given by $- 2 \gamma \lambda_{\bf w}$ and $- \gamma [\lambda_{\bf w} - \lambda_{\bf v_j}]$, where $\lambda_{\bf w}$ and $\lambda_{\bf v_j}, \forall j=\overline{1,n-1}$ are the $n$ eigenvalues of ${\bf EC}$ (noting first that ${\cal B}_{\bf w}=\{ {\bf w}, {\bf v_1},...{\bf v_{n-1}} \}$, the completion of ${\bf w}$ to a basis of eigenvectors of ${\bf EC}$, orthogonal with respect to the dot product $\langle \cdot,\cdot \rangle_{\bf C}$, also forms an eigenvector basis for the Jacobian). We concluded that, if ${\bf EC}$ has a unique largest eigenvalue, then a normalized eigenvector ${\bf w}$ is a local hyperbolic attracting equilibrium for (\ref{mothersys}) iff it corresponds to the maximal eigenvalue of ${\bf EC}$. If ${\bf EC}$ has a multiple largest eigenvalue, the system will exhibit a set of nonisolated, neutrally attracting equilibria (all normalized eigenvectors spanning the principal eigenspace, in this case of dimension $\geq 2$). Some of the computational details are summarized in Appendix 1 (e.g., a description of the attraction basins, supporting the absence of cycles in the phase space) and further expanded in our previous work~\cite{radulescu2009hebbian,radulescu2010}.

Since the nature and position of the equilibria depends on the spectral properties of ${\bf EC}$, the next task is, naturally, to study the spectral changes of ${\bf EC}$ when perturbing the system by increasing the transmission inspecificity. In our previous work on the model, we found that the effects of perturbations on the system's dynamics can depend very strongly on the characteristics of the input distribution (correlation sign, degree of bias). In our first study we only considered learning of positively correlated $n$-dimensional input distributions, and found a smooth degrading of the learning outcome with increasing error, but no sudden changes in dynamics~\cite{radulescu2009hebbian}. In our second study, we discovered that negatively correlated inputs can induce a bifurcation (stability swap of equilibria, through a critical stage) when increasing the error, even in as simple as a two-dimensional system; this bifurcation only occurred, however, in the case of unbiased inputs~\cite{radulescu2010}.

Here, we want to extend this work and investigate the effects of cross-talk in higher dimensional networks, when learning a variety of classes of input distributions, both biased and unbiased. More precisely, we will consider as potential covariance matrices all combinations of the form:

\begin{equation}
{\bf C}=\left[\begin{array}{cccc}
v+\delta_1& \pm c&\cdots & \pm c\\
 \pm c&v+\delta_2&\cdots & \pm c\\
\vdots & &\ddots &\vdots \\
 \pm c& \pm c&\cdots &v+\delta_n\\
\end{array}\right]
\label{covariancegeneral}
\end{equation}

\noindent where we can assume without loss of generality that $\delta_1 \geq \delta_2 \geq \hdots \geq \delta_n \geq 0$. For any $k \leq n$, we will say that the input has bias loss of order $k$ if $\delta_1 = \hdots = \delta_k$. We hypothesize that, even though the background covariance $\pm c$ is taken for simplicity to be uniform in absolute value, the inspecific learning rule will lead to interesting dynamics, in particular when the inputs exhibit some degree of cross-correlation.

Since our analysis will focus on symmetric matrices ${\bf C}$ with possible off-diagonal elements, we have to first ask whether/when such a matrix can constitute the covariance matrix of a distribution of $n$-dimensional vectors. While establishing equivalent conditions may be difficult even for small dimensions~\cite{vasudeva1998negative}, a simple sufficient criterion valid for any dimension is diagonal dominance. It is known that a symmetric diagonally dominant matrix with real, non-negative diagonal entries is positive semi-definite, hence implicitly a covariance matrix, from the finite-dimensional case of the spectral theorem\footnote{If ${\bf X}$ is an $n \times 1$ column vector-valued random variable whose covariance matrix is the $n \times n$ identity matrix. Then $\text{cov} \; (\sqrt{\bf C}{\bf X}) = \sqrt{\bf C} \; \text{cov}({\bf X}) \; \sqrt{\bf C} = {\bf C}$.}. If we are willing to impose $v+\delta_n > (n-1)\vert c \vert$ as a (biologically plausible) upper bound on how large the input cross-correlations $\lvert c \rvert$ can be with respect to the auto-correlations $v$, diagonal dominance clearly follows, and ${\bf C}$ is thus automatically guaranteed to be a covariance matrix. An interesting direction would be to interpret biologically the significance of an $n$-dimensional input distribution with negative correlations~\cite{durrant1998negative}; this question is, however, beyond the scope of this paper.

\section{Classes of inputs and bias effects on 3-dimensional dynamics}
\label{3dim}

We study here how input patterns affect the effects of inspecificity in driving the dynamics of a 3-dimensional network -- the lowest dimension for which the question applies, and which captures the essence of this behavior even in higher-dimensional systems. In this section, we will inspect all combinatorial possibilities of cross-correlation sign and auto-correlation bias, and determine the effect of increasing error on the dynamics in each case. In the next section, we will support with some rigorous proofs the main results obtained here through numerical simulations (we used the Matlab software, version 7.2.1).

\subsection{Input covariance patterns}
\label{input3dim}

We studied separately all combinatorial possibilities for the input statistics, with uniform absolute value covariance; in other words, we considered covariance matrices of the form:\\

\begin{equation}
{\bf C}=\left[\begin{array}{cccc}
v+\delta_1 & \pm c & \pm c \\
\pm c & v+\delta_2 & \pm c \\
\pm c & \pm c & v \\
\end{array}\right]
\label{covariance3d}
\end{equation}

\noindent where, as before, $v > 2\vert c \vert$, and $\delta_1 \geq \delta_2 \geq 0$ (i.e., allowing bias of any order).

\begin{figure}[h!]
\begin{center}
\includegraphics[scale=0.22]{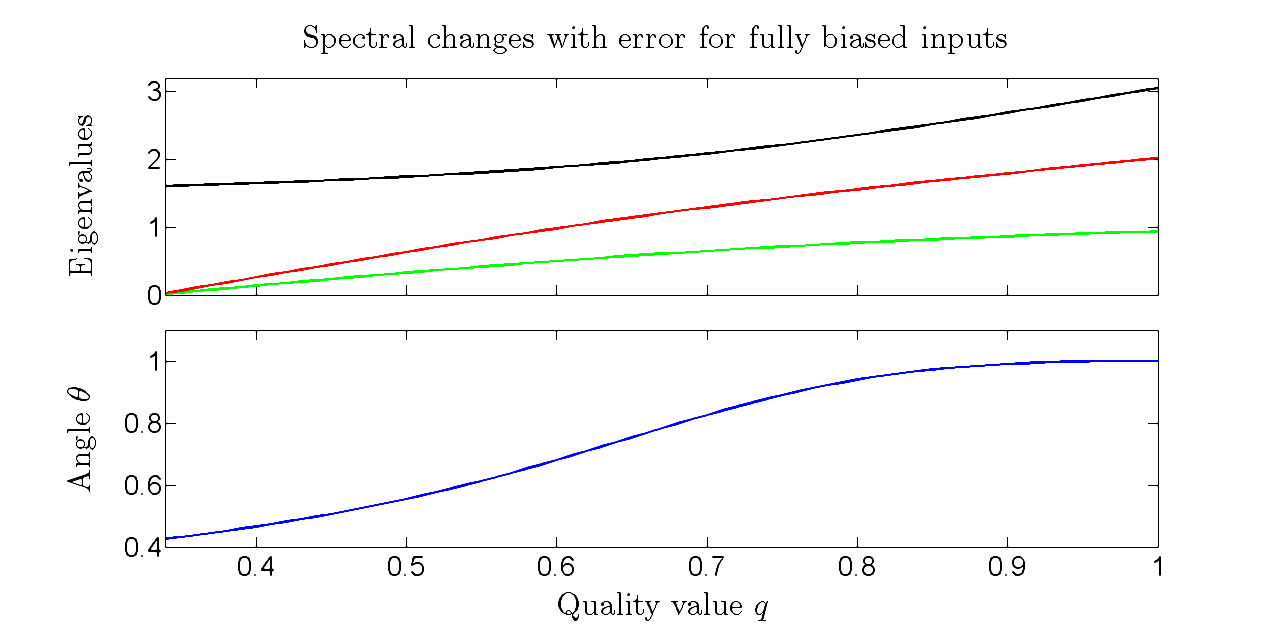}
\includegraphics[scale=0.22]{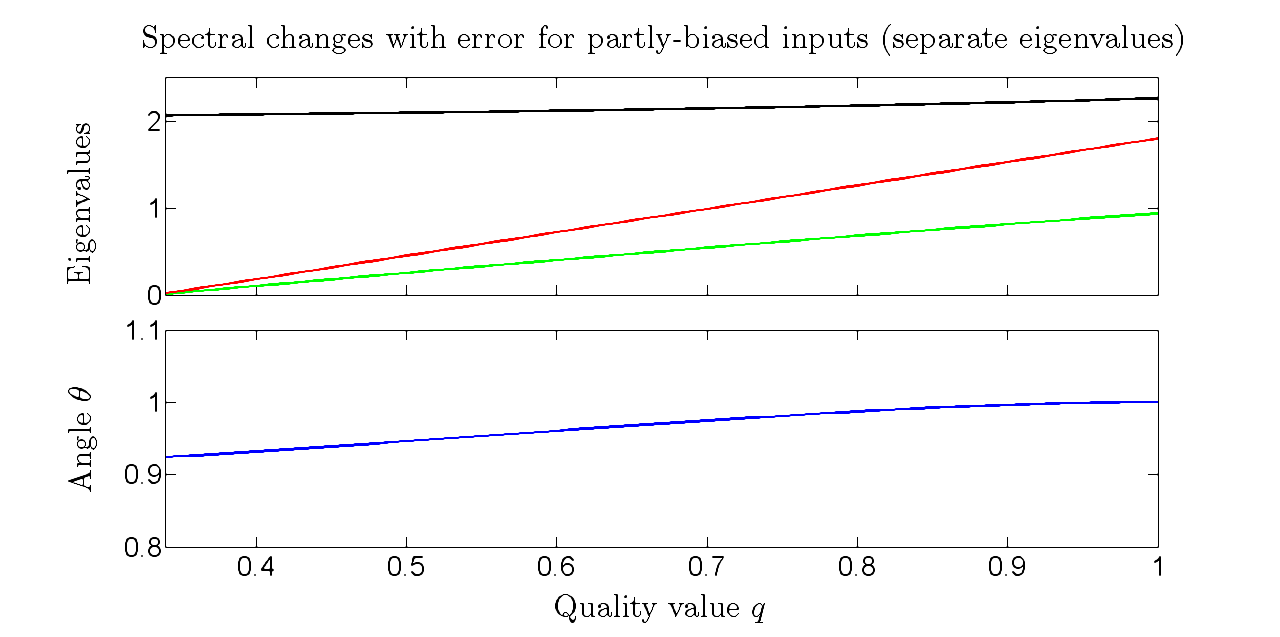}\\
\includegraphics[scale=0.22]{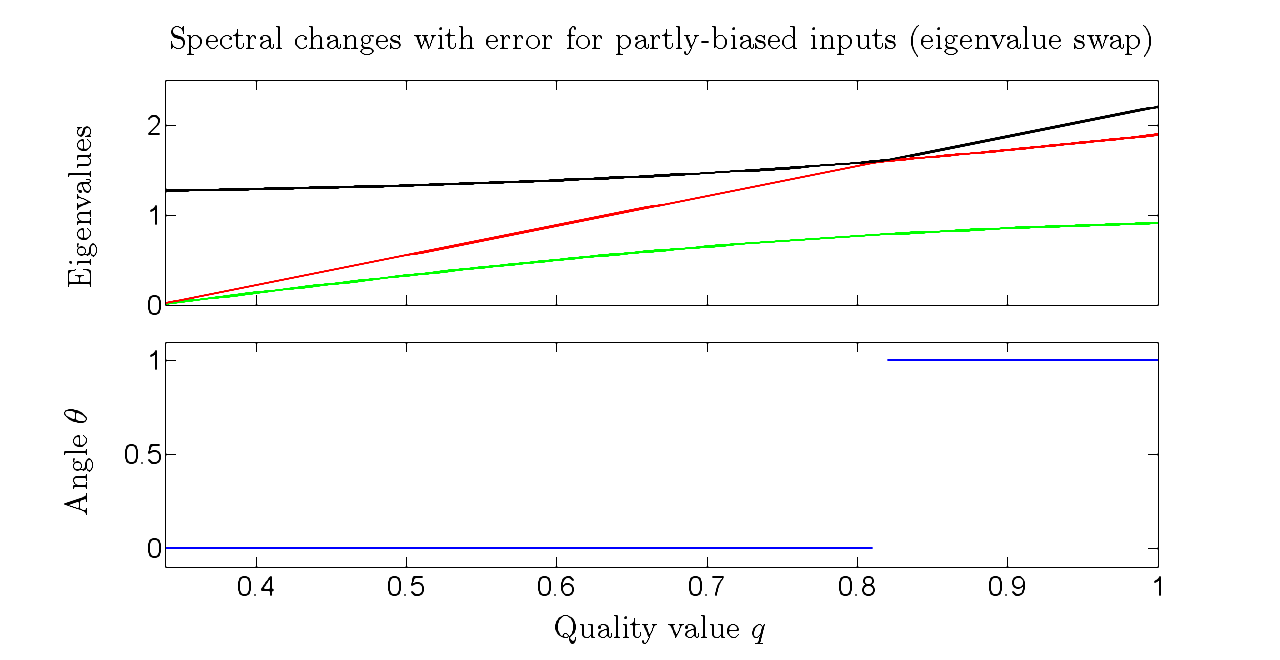}
\includegraphics[scale=0.22]{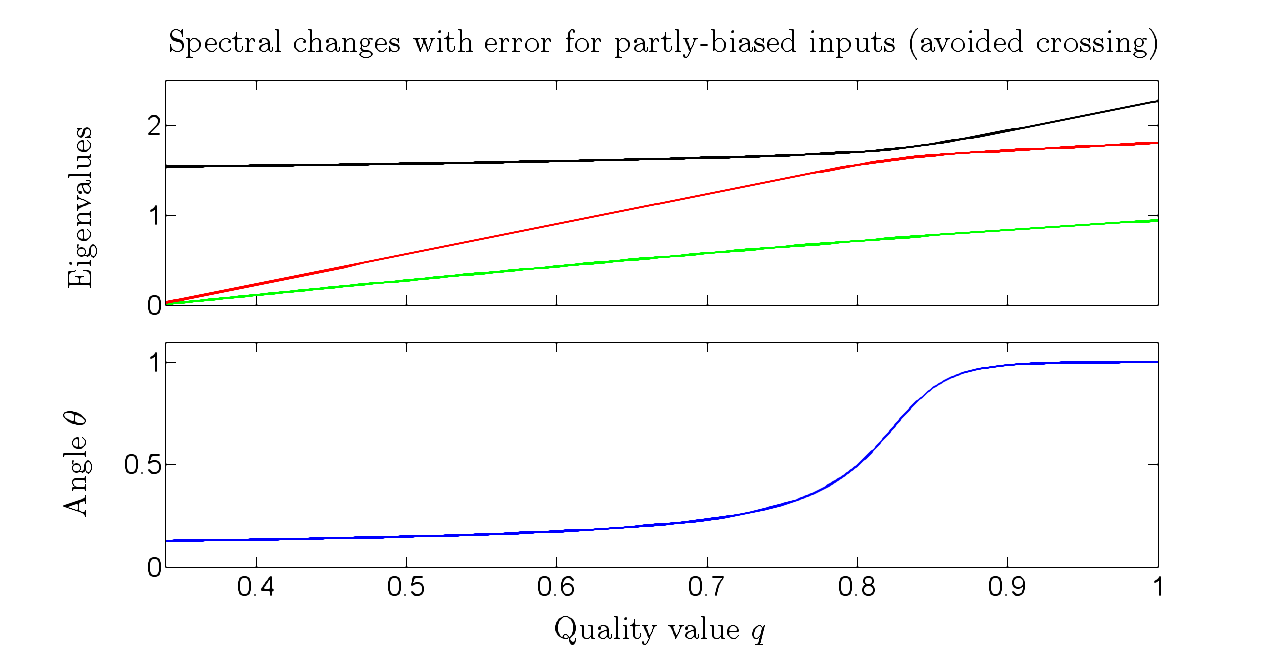}\\
\includegraphics[scale=0.22]{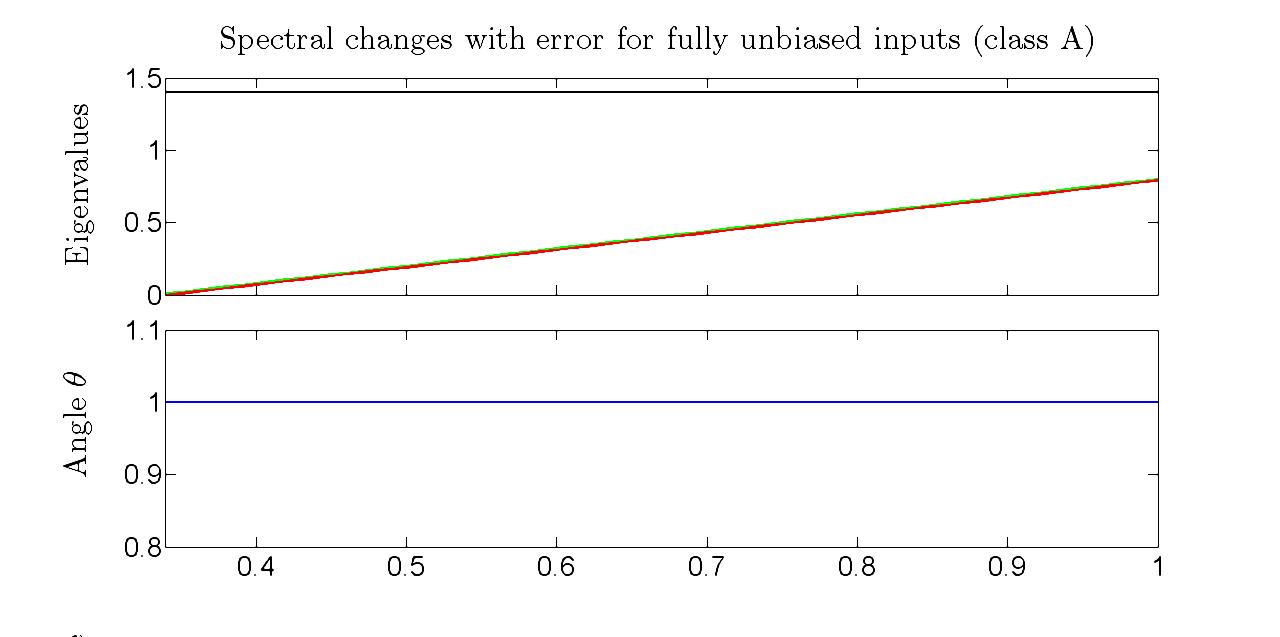}
\includegraphics[scale=0.22]{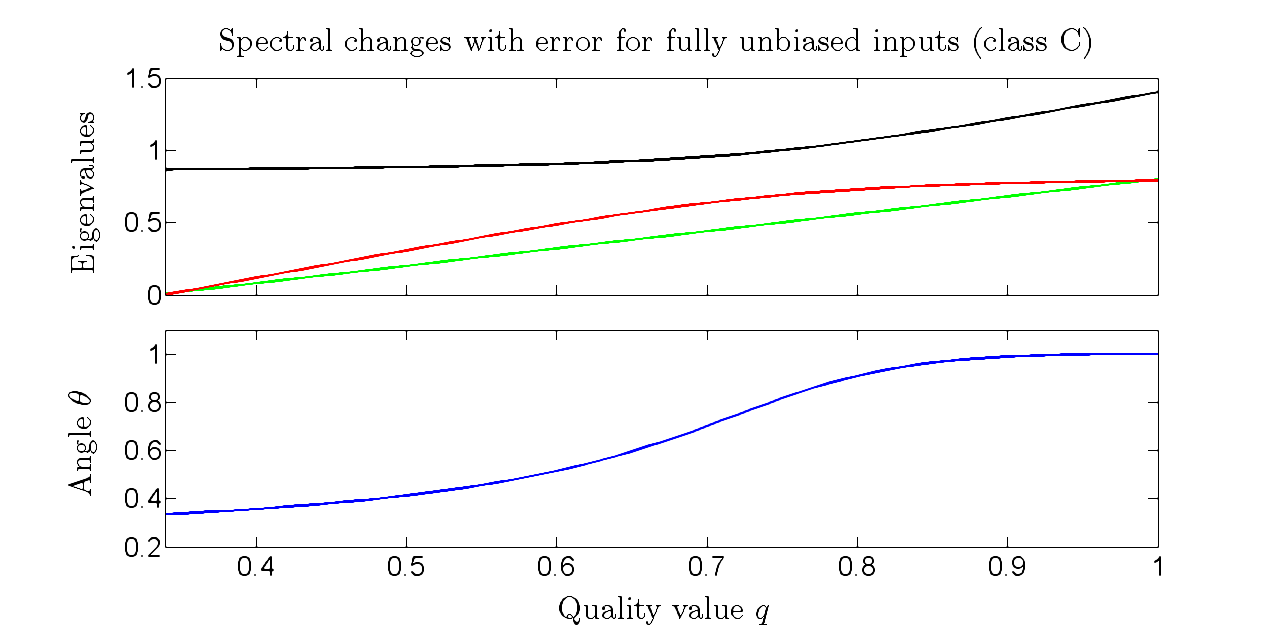}\\
\includegraphics[scale=0.22]{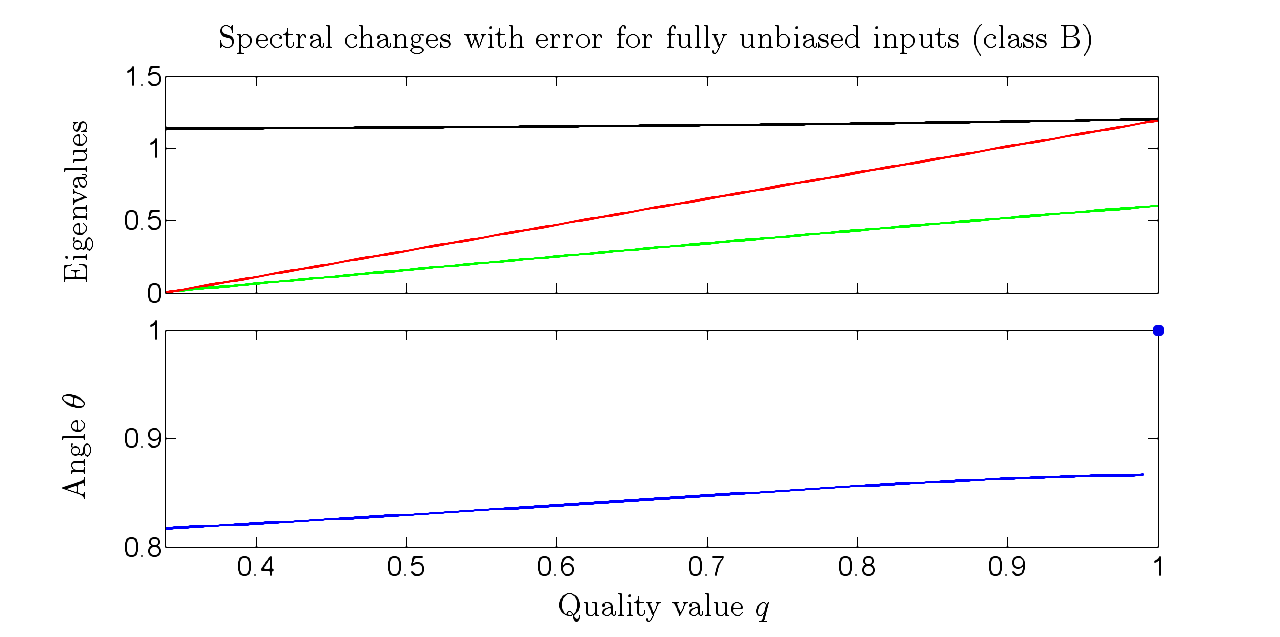}
\includegraphics[scale=0.22]{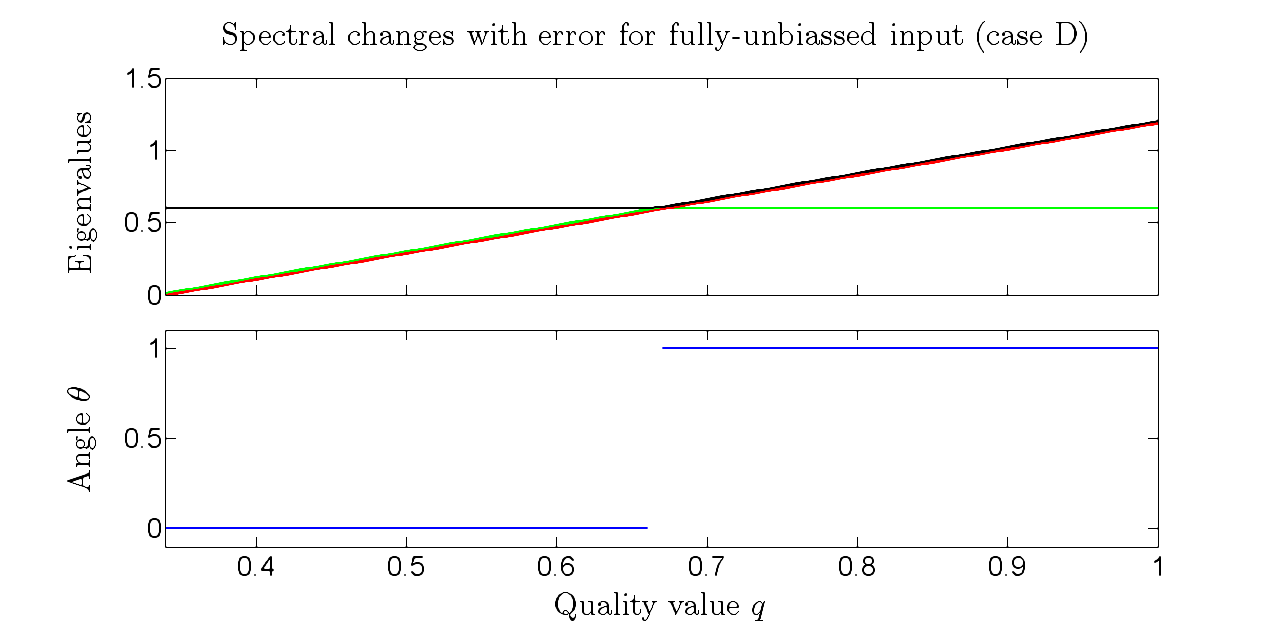}
\end{center}

\caption{\small \emph{{\bf Spectral changes induced by increasing inspecificity, for various inputs schemes.} In all panels we show, with respect to the quality $q=1-2\epsilon$: the evolution of the eigenvalues, with color-coding black for largest eigenvalue, red for the second largest and green for the lowest (top subplot); the angle between the inspecific stable vector and the correct attracting direction(s) (bottom subplot). In all panels, $v=1$, $\lvert c \rvert = 0.2$. The classification is as follows: {\bf A.} For fully biased inputs ($\delta_1 =2$, $\delta_2=1$), the three eigenvalues remain separated. For partly-biased inputs ($\delta_1=\delta_2=1$), there are three cases, depending on the number of negative cross-correlations and on their placement: the leading eigenvalues can remain separated ({\bf B}), they can cross at a critical values of $q=q^*$ ({\bf C}), or they can approach significantly for some value of $q$, but ``avoid'' crossing ({\bf D}). For fully unbiased inputs, we found four cases, classified simply by the number of negative off-diagonal cross-correlations (and not by their geometry): all positive cross-correlations --- leading eigenvalues remain separated ({\bf E}); one negative cross-correlation --- leading eigenvalues only coincide at $q=1$, and immediately separate ({\bf F}); two negative cross-correlations --- leading eigenvalues may approach each other, in an avoided crossing of magnitude depending on parameters, but remain separated ({\bf G}); all negative cross-correlations --- leading eigenvalues coincide on a whole interval, as quality depreciates from $q=1$ to a critical value. In this case, the system has a curve of half-neutral attractors, which persists until $q$ reaches the critical value, when a different, orthogonal, eigenvector takes over as stable direction.}}
\label{spectrum}
\end{figure}

Let's first note that, based on the number of negative upper-diagonal entries of $C$, we distinguish 4 combinatorial classes: (A) all positive covariance (one configuration), (B) one negative entry (3 configurations), (C) two negative entries (3 configurations), (D) and all negative entries (one configuration). We will study the spectra of the inspecific matrices ${\bf EC}$, and the differences that occur in these when considering different classes of ${\bf C}$, as well as different degrees of bias: from fully biased ($\delta_1 > \delta_2 > 0$) to partly biased ($\delta_1 = \delta_2 > 0$) to fully unbiased ($\delta_1 = \delta_2 = 0$). In this section, we illustrate the behavior of the eigenvalues of ${\bf EC}$ as the quality $q$ is changing in the interval $(1/3, 1]$ (representing quality higher than error).\\

\begin{figure}[h!]
\begin{center}
\includegraphics[scale=0.31]{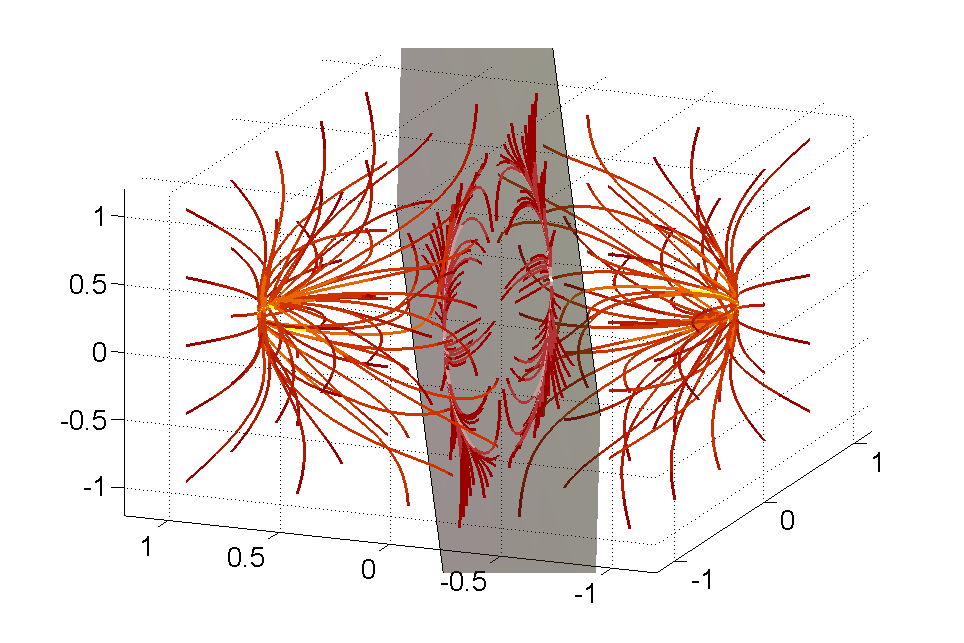}
\includegraphics[scale=0.31]{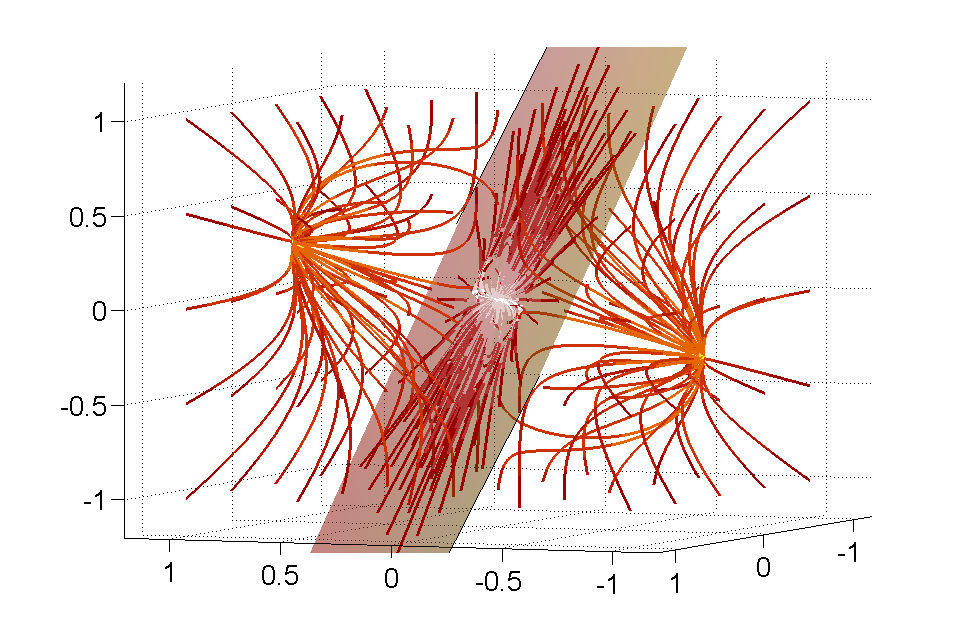}
\end{center}
\caption{\small \emph{{\bf Phase space trajectories for fully biased inputs.} {\bf A.} In the absence of error, the system converges generically to the two normalized vectors in the principal direction ${\bf w}_{\bf C}$ of the covariance matrix ${\bf C}$. The attraction basins are separated by the subspace $\langle {\bf w}, {\bf w_{C}} \rangle = 0$ (the shaded plane). {\bf B.} For error $\epsilon=0.2$, the system converges generically to the two normalized vectors in the principal direction ${\bf w}_{\bf EC}$ of the modified covariance matrix ${\bf EC}$. The attraction basins are separated by the subspace $\langle {\bf w}, {\bf w_{EC}} \rangle = 0$ (the shaded plane). Parameters used: $v=1$, $c=0.2$, $\delta_1=2$, $\delta_2=1$. Color coding: trajectories evolve in time from darker towards lighter shades.}}
\label{phsp_full_bias}
\end{figure}

\noindent {\bf For fully biased inputs} ($\delta_1 > \delta_2 > 0$), the behavior is indistinguishable between classes\footnote{Since the spectra depend qualitatively on all parameter values, we present here the results of a numerical investigation, rather than an rigorous analytical study, which would be extremely cumbersome. In contrast, we will later prefer an analytical approach to the classification in fully unbiased case, where the computations become more tractable.}: the largest eigenvalue remains separated from the second largest for the whole range of $q$ (as shown for one example in Figure \ref{spectrum}a), determining the eigenvector to gradually drift from the direction of the principal component of ${\bf C}$ (blue curve in Figures \ref{spectrum}a). For any value of $q$, the system has two hyperbolically attracting equilibria (the normalized principal eigenvectors of ${\bf EC}$, whose basins are separated by an invariant plane). In Figure \ref{phsp_full_bias} we show the evolution of a set of trajectories, to illustrate convergence to the two attractors in the phase space, as well the dynamics within the separating plane. (We will encounter similar behavior for other classes of inputs as well, for which we will refer to the same Figure, since the same phase space evolution remains a qualitatively accurate depiction.)\\

\noindent {\bf For loss of bias of order one} ($\delta_1 = \delta_2 > 0$), we distinguish three possible types of behavior.

\vspace{2mm}
\noindent {\bf \emph{Separated leading eigenvalues}} (behavior ``typical'' to class A, as illustrated in Figure \ref{spectrum}b). This corresponds to a slow depreciation of the learned vector as $q$ decreases (blue curve in Figure \ref{spectrum}b). The phase space behavior resembles qualitatively that for unbiased inputs, illustrated in Figure \ref{phsp_full_bias}.

\begin{figure}[h!]
\begin{center}
\includegraphics[scale=0.31]{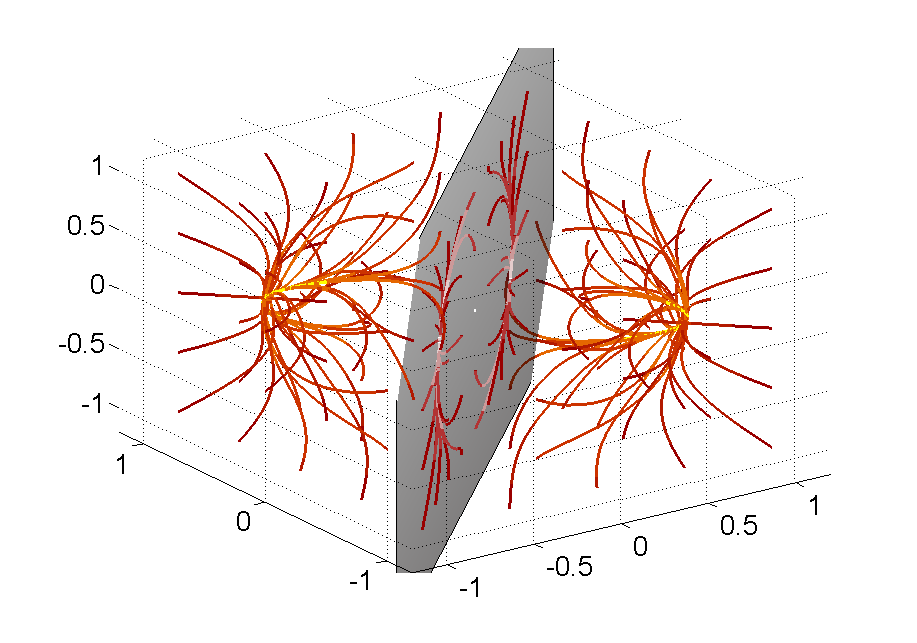}
\includegraphics[scale=0.31]{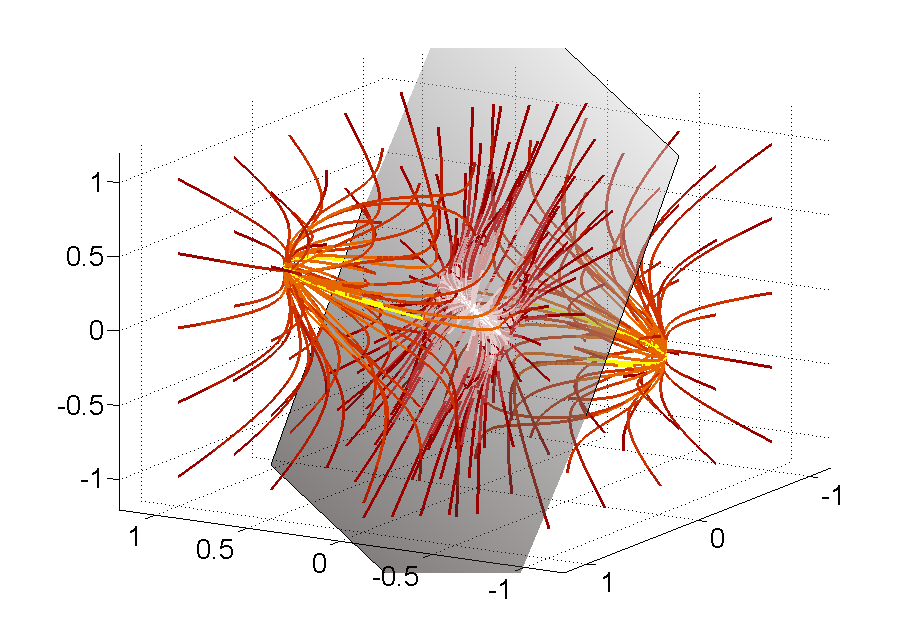}\\
\includegraphics[scale=0.31]{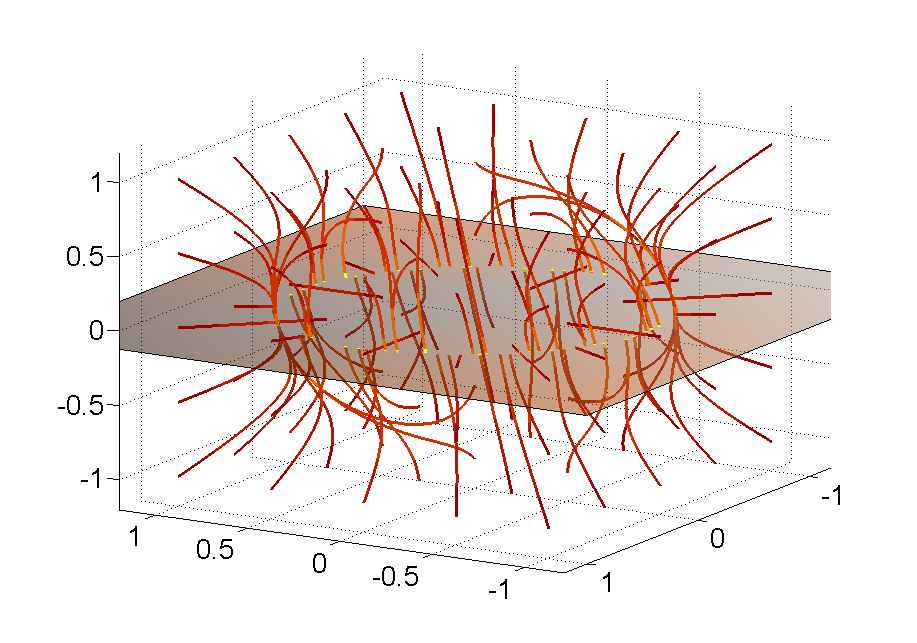}
\end{center}
\caption{\small \emph{{\bf Bifurcation in attractor dynamics for partly biased inputs, all negative cross-correlations.} {\bf A.} For small error, the attractors (the two normalized principal eigenvectors of ${\bf EC}$) don't differ too much from the correct attractors (the two normalized principal eigenvectors of ${\bf EC}$). The attraction basins are separated by the subspace $\langle {\bf w}, {\bf w_{C}} \rangle = 0$ (the shaded plane). {\bf B.} For critical error $\epsilon=\frac{-c}{v+\delta-c}$, the system exhibits an ellipse of neutrally-stable equilibria (yellow curve contained in the shaded plane). {\bf C.} For error past the critical value, the attractors have moved significantly far from the correct positions. Parameters used: $v=1$, $c=0.2$, $\delta=\delta_1=\delta_2=1$. Color coding: trajectories evolve in time from darker towards lighter shades.}}
\label{phsp_part_bias}
\end{figure}

\vspace{2mm}
\noindent {\bf \emph{Crossing of leading eigenvalues}} (behavior ``typical'' to class D, as illustrated in Figure \ref{spectrum}c and further discussed in Section \ref{ndim}), which produces a sudden swap of the attractors from one eigendirection to another, orthogonal, one (phenomenon we have described previously in a 2-dimensional model~\cite{radulescu2010}). This corresponds to a crash in learning at a critical value of the quality $q^*$ (which depends on parameters as $q^*=\frac{v+\delta+c}{v+\delta-c}$). Low inspecificity  ($q>q^*$) has in fact no effect on learning in this case: although the leading eigenvalue changes, the principal direction remains the same, so the system will converge generically to the same outcome as in the absence of error. This may seem like a very desirable input distribution to learn in the presence of inpecificity; however, one has to keep in mind that, if the cross-correlations are small in absolute value $\lvert c \rvert$, then $q^*$ will be very close to 1. Such perfect learning will therefore only happen when inspecificity is almost insignificantly small. The more disturbing this becomes, when we recall that at the end of the ``good'' interval lies the bifurcation, crashing the equilibria to a completely irrelevant direction; so any fault of the system in the direction of slightly miscalculating the limits for the permissible error, would have dire consequences. If the network does not have an additional, good estimator of its degree of inspecificity, it may not only learn an irrelevant outcome, but also have no knowledge of it.  In Figure \ref{phsp_part_bias}, we represent three phase space plots: before, at and after the bifurcation point. While Figures \ref{phsp_part_bias}a and \ref{phsp_part_bias}c illustrate the typical phase space with two hyperbolically stable equilibria (one representing accurate, error-free learning, the other -- inaccurate learning for a post-critical error), the bifurcation phase space is qualitatively different: the system has no hyperbolic attractors, but rather a closed curve (ellipse) of half-stable equilibria (neutral along the direction of the curve). Clearly, the outcome of learning is in this case extremely dependent on the initial conditions (although, as commented in previous work~\cite{radulescu2010}, the stochastic version of the system will rather have noise-driven stationary solutions that drift \emph{around} this attracting ellipse). If this phase-plane dynamics were specific only to this bifurcation, one may find it justified to overlook its occurrence in the context of generic dynamics. However, this is not the case; as shown below, there are classes of inputs for which such a attracting-ellipse slice represents the natural state of the system, and persists for an entire inspecificity range.

\vspace{2mm}
\noindent {\bf \emph{``Avoided crossing'' of the eigenvalues}} (a hybrid behavior observed in mixed cases from classes B and C, in which the eigenvalues approach, without actually crossing, at a value $q=q^*$, which depends on all parameter values). While the principal eigenvectors never swap in this case, the learning has a significantly rapid depreciation around $q^*$ (see blue curve in Figure \ref{spectrum}d).\\

\noindent {\bf For bias loss of order two}, the computations are much simplified by the absence of bias, so we can carry out analytically a complete classification. The result is presented concisely in Theorem \ref{classification}, explained in more detail in the proof of the theorem, then further interpreted for the remainder of the section.

\begin{thm}
\label{classification}
For order two input bias $\delta_1 = \delta_2 = 0$, the dynamic behavior  of the system is classified by the classification of the input covariance sign, {\bf A}-{\bf D}.
\end{thm}

\proof{For order two input bias, all classes {\bf A}-{\bf D} can be generated from three symbolic structures:

\begin{equation}
{\bf C}_1=\left[\begin{array}{cccc}
v & c & c \\
c & v & c \\
c & c & v \\
\end{array}\right]
\;
,
\;
{\bf C}_2=\left[\begin{array}{cccc}
v & -c & c \\
-c & v & c \\
c & c & v \\
\end{array}\right]
\;
\text{ and}
\;
{\bf C}_3=\left[\begin{array}{cccc}
v & c & -c \\
c & v & c \\
-c & c & v \\
\end{array}\right]
\nonumber
\end{equation}

\noindent Class {\bf A} represents Structure ${\bf C}_1$ with $c>0$, and Class {\bf D}, represents Structure ${\bf C}_1$ with $c<0$. Class {\bf B} can be obtained from Structures ${\bf C}_2$ and ${\bf C}_3$ with $c>0$, while Class {\bf C} can be obtained from Structures ${\bf C}_2$ and ${\bf C}_3$ with $c<0$.

Computing directly the spectrum for ${\bf C}_1$, we get one simple eigenvalue $\xi_1=v+2c$ (whose eigenvector is also error-independent) and one double eigenvalue $\xi_2=(1-3e)(v-c)$. If $c>0$ (Class {\bf A}), $\xi_1$ always dominates (Figure \ref{spectrum}e). If $c<0$ (Class {\bf D}), the double eigenvalue $\xi_2=(1-3e)(v-c)$ takes over for error smaller than the critical value $\ds \epsilon < \frac{-c}{v-c}$ (Figure \ref{spectrum}h).

Also by direct computation, one notices that ${bf C}_1$ and ${\bf C}_2$ have the same spectral decomposition. One eigenvalue is given by $\xi_1=(1-3e)(v+c)$, while the other two $\xi_2 \geq \xi_3$ are the roots of the quadratic polynomial $P(X) = X^2+(c-2v-5ec+3ev)X+(6ec^2-cv-3ev^2-2c^2+v^2+3ecv)$. It is easy to see that $P(\xi_1)=-8ec(1-3e)(v+c)$. If $c>0$ (Class {\bf B}), then $P(\xi_1)<0$, hence $\xi_2 \geq \xi_1$, with equality at $\epsilon=0$, and $\xi_1 \geq \xi_3$, with equality when $\epsilon=1/3$ (Figure \ref{spectrum}g). If $c<0$ (Class {\bf C}), then $P(\xi_1)>0$ and $\xi_1 < (\xi_2+\xi_3)/2$, hence $\xi_1 \leq \xi_2 < \xi_3$, with equality when $\epsilon=0$ and $\epsilon=1/3$ (Figure \ref{spectrum}f).
}

\noindent The theorem allows us some immediate class-specific interpretations in the context of phase space dynamics and learning.\\

\begin{figure}[h!]
\begin{center}
\includegraphics[scale=0.31]{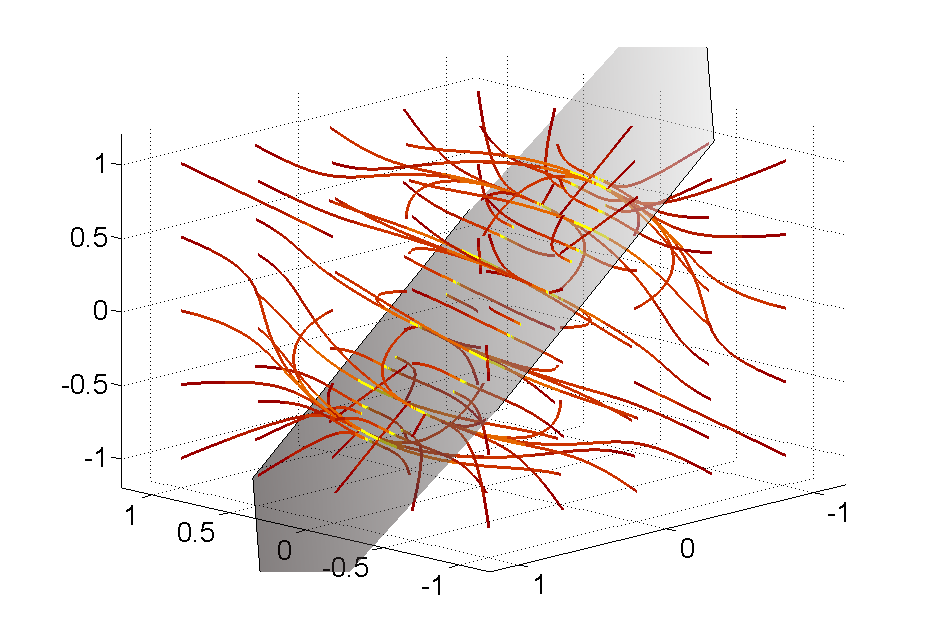}
\includegraphics[scale=0.31]{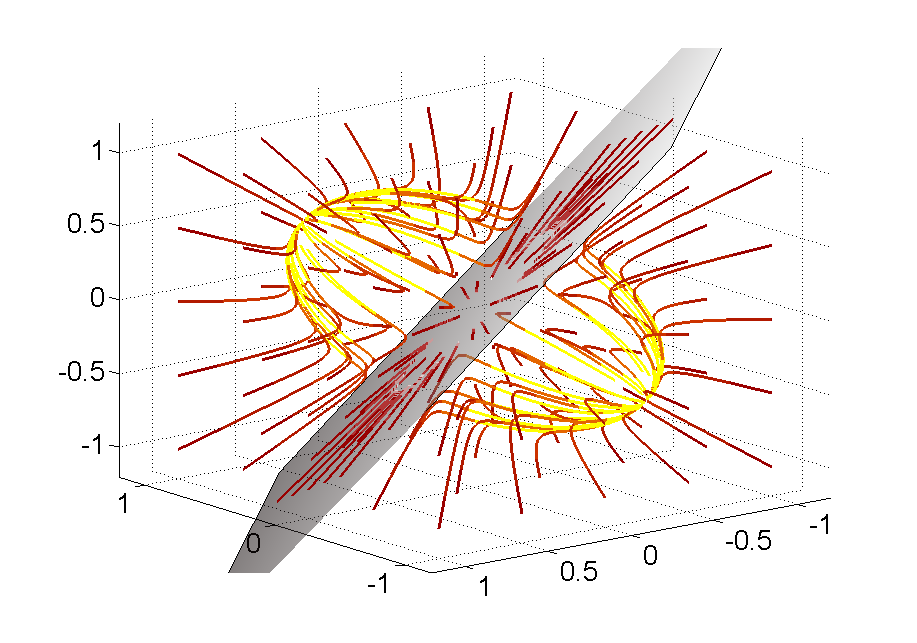}
\end{center}
\caption{\small \emph{{\bf Bifurcation in attractor dynamics for partly biased inputs, all negative cross-correlations.} {\bf A.} For small error, the system has an ellipse of neutrally-stable equilibria (yellow curve). This ellipse is stable, in the sense that it persists for a whole interval of errors, from $\epsilon=0$ until $\epsilon=\frac{-c}{v-c}$. {\bf B.} For error past the critical value, the ellipse is destroyed, but the new attractors are significantly far from the plane of the ellipse. Parameters used: $v=1$, $c=0.2$, $\delta_1=2=\delta_2=0$. Color coding: trajectories evolve in time from darker towards lighter shades.}}
\label{phsp_full_unbias}
\end{figure}

\vspace{2mm}
\noindent {\bf \emph{Class A.}} The leading eigenvalue is constant, and always separated from the second (double) eigenvalue. Moreover, the principal component of ${\bf EC}$ does not change when the error increases, so in this case the learning is fully accurate for any degree of inspecificity (Figure \ref{spectrum}e). This is a class of input statistics which is completely error-proof.

\vspace{2mm}
\noindent {\bf \emph{Class B.}} This falls within the typical case of separated leading eigenvalues, where the system learns, for any error value, the leading eigendirection of ${\bf EC}$ (which degrades smoothly from the principal component of ${\bf C}$; see Figure \ref{spectrum}g, and Figure \ref{phsp_full_bias}). Depending on parameters, the eigenvalue curves with respect to $q$ may exhibit a significant point of minimal separation (see ``avoided crossing''), where the learning outcome (leading eigenvector of ${\bf EC}$) deteriorates very fast.

\vspace{2mm}
\noindent {\bf \emph{Class C.}} In the error-free case, the matrix ${\bf C}$ has a double leading eigenvalue, and the system has a whole closed curve of neutrally attracting equilibria (in the eigenplane spanned by the corresponding eigenvectors). When error is introduced, the two leading eigenvalues segregate, and one of the eigenvectors takes over, which determines an immediate complete switch in the learning outcome. In this case, even the smallest degree of inspecificity leads to favoring one specific direction, slightly detaching off the plane where the ``real'' equilibria are contained (notice that the cosine of the accuracy angle, represented by the blue curve in Figure \ref{spectrum}f, does not fall too far off the perfect value $\cos(\theta)=1$). We may interpret this as the error helping the system ``make up its mind'' in the presence of too much ambiguity in the input statistics.

\vspace{2mm}
\noindent {\bf \emph{Class D.}} In the error-free case, the matrix ${\bf C}$ has a double leading eigenvalue, and the system has again a whole curve of neutral equilibria, contained in the corresponding eigenplane. When subject to errors up to a critical value $q^*=\frac{v+c}{v-c}$, the leading eigenvalues change, but remain equal; furthermore, the subspace spanned by the two corresponding eigenvectors remains unchanged, hence the learning process retains the original ambiguity. Past the critical error value, the eigenvalues swap, and the eigendirection of the new leading eigenvalue (of multiplicity one) is orthogonal to the previous plane (Figure \ref{spectrum}h). In other words, past the critical error value, the system will learn, but with such low accuracy, that the result of learning is useless.

\section{An analytical application in higher dimensions}
\label{ndim}

We will work out an analytical computation which suggests that the cases described in Section \ref{3dim} can be extended to classify the behavior of the higher dimensional inspecific Oja system. For simplicity, we consider only one application: for negatively cross-correlated inputs:

\begin{equation}
{\bf C} = \left[\begin{array}{cccc}
v+\delta_1 & -c & \cdots & -c\\
-c & v+\delta_2 & \cdots & -c\\
\vdots & & \ddots & \vdots \\
-c & -c & \cdots & v+\delta_n\\
\end{array}\right]
\label{covariance}
\end{equation}

\noindent Although this is perhaps the least biologically sound case, we feel that it is mathematically the most interesting, and describes the opposite scenario from the case of all positively cross-correlated inputs (which is mathematically the least interesting). As suggested by the numerical computations in 3 dimensions, covariance matrices that exhibit other positive/negative patterns of cross-correlations are expected to produce hybrid dynamics between these two extreme ends. These dynamics will depend not only on the number of negative correlations, but also on their distribution within the covariance matrix. A random matrix analysis may be able to classify behavior for all input patterns, but this is not within the scope of this study. In this section, we only present the main analytical results we obtained for our application; proofs of the statements and additional comments can be found in Appendix 2.\\

\noindent {\bf Fully biased case.} We first consider the covariance biases $\delta_j$'s to be distinct:  $\delta_1 > \delta_2 > \hdots > \delta_{n-1} > \delta_n = 0$. We want to study the eigenvalues and eigenspaces of the modified covariance matrix ${\bf EC}$. The characteristic polynomial of ${\bf EC}$ can be expressed as (see details in Appendix 2):\\

$\Delta(\lambda)=\det({\bf EC}-\lambda {\bf I})=\left \lvert \begin{array}{cccc}
X_1(\lambda)&f_2&\cdots &f_n\\
f_1&X_2(\lambda)&\cdots &f_n\\
\vdots & &\ddots &\vdots \\
f_1&f_2&\cdots &X_n(\lambda)\\
\end{array}\right \rvert$\\

\noindent where for all $j=\overline{1,n}$, we called $f_j=\epsilon(v+\delta_j-c)+c$ and $X_j(\lambda)=q(v+\delta_j-c)+c -\lambda$.\\

We consider $\lambda_j=(q-\epsilon)(v+\delta_j-c)$; clearly: $\lambda_1 >\lambda_2> ... >\lambda_n$. In Appendix 2, we show how these values can be used to partition the real line and separate the roots of $\Delta$. In particular, we prove the following:

\begin{prop}
In the biased case $\delta_1>\delta_2> \hdots >\delta_n$, the matrix ${\bf EC}$ has $n$ real distinct eigenvalues $\xi_1>\xi_2> \hdots > \xi_n$.
\end{prop}

We can define, as in the 2 and 3-dimensional applications, the ``critical'' error values, for which $f_j(\epsilon_j^*)=0, \; \forall j \in \overline{1,n}$:

\begin{equation}
\epsilon_j^*= \frac{-c}{v+\delta_j-c}
\end{equation}

\noindent so that $0<\epsilon_1^*<\epsilon_2^*< \hdots < \epsilon_n^*$ (since $\delta_1>\delta_2> \hdots > \delta_n$). Clearly, for all $j \in \overline{1,n}$, we have $f_j >0$ iff $\epsilon > \epsilon_j^*$. As $\epsilon$ increases from $0$ to $1/n$, it traverses the values $\epsilon=\epsilon_j^*$. When $\epsilon$ is in the intervals between two consecutive critical values $\epsilon_j^*$, each two consecutive roots of $\Delta$ are separated by at least one $\lambda_j$. When $\epsilon$ reaches each critical value $\epsilon_j$, the root $\xi_j$ crosses from one interval to another through the stage $\xi_j=\lambda_j$.

\begin{figure}[h!]
\begin{center}
\includegraphics[scale=0.52]{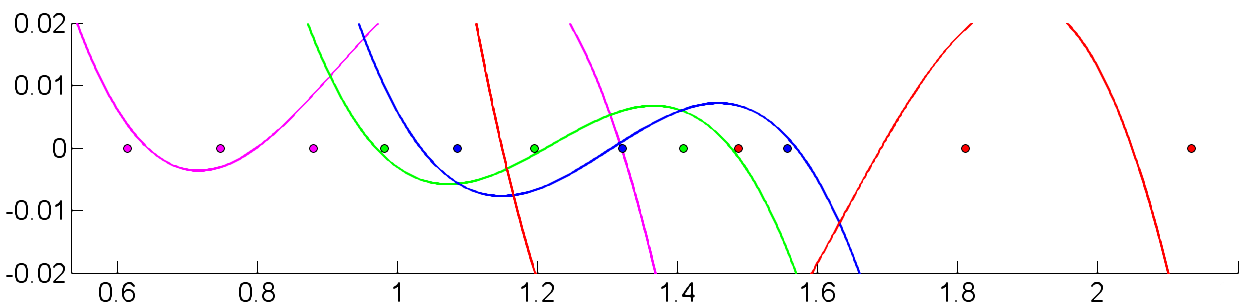}
\caption{{\small {\bf A simple example of how the characteristic polynomial $\Delta$ of ${\bf EC}$ and its roots change as the quality $q$ decreases}, for dimension $n=3$ and fixed parameters $v=1$, $c=-0.2$, $\delta_j=j/3$, so that $\epsilon_1^* \sim 0.091$, $\epsilon_2^* \sim 0.107$, $\epsilon_3^* \sim 0.130$, for $j \in \overline{1,3}$. Each different color represents a different values of $q$: $q=0.98$ (red), $q=0.805$ (blue), $q=0.76$ (green), $q=0.6$ (pink). The continuous curves correspond to the graph of the polynomial for different $q$'s, and the bullets represent (along the x-axis) the points $\lambda_j=(q-\epsilon)(v+\delta_j-c)$, for $j=\overline{1,3}$. The figure shows how the order of the position of the roots of $\Delta$ changes with respect to the points of the partition $\lambda_3<\lambda_2<\lambda_1$ (which in turn travel down the axis as $q$ decreases). For $q=0.98$ (i.e., $\epsilon=0.01<\epsilon_1^*$), $\lambda_1>\xi_1>\lambda_2>\xi_2>\lambda_3>\xi_3$. For $q=0.805$ (where $\epsilon=0.0975\in [\epsilon_1^*,\epsilon_2^*]$), $\xi_1>\lambda_1>\lambda_2>\xi_2>\lambda_3>\xi_3$.  For $q=0.76$ (where $\epsilon=0.12 \in [\epsilon_2^*,\epsilon_3^*]$), $\xi_1>\lambda_1>\xi_2>\lambda_2>\lambda_3>\xi_3$.  For $q=0.6$ (where $\epsilon=0.2>\epsilon_3^*$), $\xi_1>\lambda_1>\xi_2>\lambda_2>\xi_3>\lambda_3$.}}
\end{center}
\end{figure}

\vspace{5mm}
\noindent {\bf Losing the bias.} Suppose now that, for $j \in \overline{1,n}$, $\delta_j = \delta_{j+1}+\zeta_j$, and allow some of the $\zeta_j \to 0$; in the limit, this results in a loss of bias in the covariance matrix ${\bf C}$ ($v+\delta_j=v+\delta_{j+1}$ for some index $j$). In consequence, $\lambda_j - \lambda_{j+1} \to 0$. It follows that in the limit of $\zeta=0$ and $\xi=\lambda_1=\lambda_2$, so that the maximal eigenvalue of ${\bf EC}$ preserves its multiplicity =1. This situation changes if we introduce an \emph{order two bias loss} $\delta_1=\delta_2=\delta_3$ (i.e. if we make both $\zeta_1$ and $\zeta_2$ approach zero simultaneously). Then $\lambda_1-\lambda_2 \to 0$ and $\lambda_2-\lambda_3 \to 0$, so that, the two leading roots collide into a double root $\lambda_3=\xi_2=\lambda_2=\xi_1=\lambda_1$. This justifies the following proposition:

\begin{prop}
Suppose $\epsilon<\epsilon_1^*$. An \emph{order $k$ bias loss} of the covariance matrix ${\bf C}$ of the type $\delta_1=\hdots=\delta_k$ results in a leading eigenvalue of multiplicity $k-1$ for the modified covariance matrix ${\bf EC}$.
\end{prop}

\section{Discussion}
\label{disc}

In this study, we considered a learning network based on the classical unsupervised learning model of Oja, extended it to allow synaptic cross-talk (encoded either as inspecificity $\epsilon$, or as transmission quality $q=1-(n-1)\epsilon$) and we showed how different input patters can exacerbate, or at the contrary, efface the effects of this cross-talk on the asymptotic outcome of learning.

We made a few simplifying assumptions: we considered uniform magnitude of input cross-correlations (i.e, uniform absolute value $\lvert c \rvert$ of the off-diagonal elements of ${\bf C}$), and uniform error (the Hebbian adjustment of any weight was equally affected by error, and did not depend either on the strength of that weight or its identity). Such ``isotropicity'' seems like a reasonable basic assumption, and has been further motivated and discussed in our previous work~\cite{radulescu2009hebbian, radulescu2010}. Furthermore, it allowed us to identify other features of the input distribution, crucially consequential on the learning dynamics and outcome: the cross-correlation signs, and the input bias.

When observing the (qualitative and quantitative) effects that the presence of cross-talk can have on the system's asymptotic behavior, we noted that these can vary substantially, depending on the input second-order statistics. We found that, in specific highly unbiased cases, the cross-talk has no effect on the presence and position of the asymptotic attractors (Figure \ref{spectrum}e). In other cases, the depreciation of the asymptotic outcome with error is so slow, that small errors have virtually no effect on learning (Figure \ref{spectrum}a,b; also see Figure \ref{phsp_full_bias} for a phase space illustration).

Other significant classes of inputs, however, exhibited a sudden change of the attracting direction from an almost perfect input principal component estimator to a direction almost orthogonal to the original. This occurred either in the form of an eigenvalue swapping bifurcation in dynamics (producing the instantaneous loss of learning accuracy at a critical error value; see Figure \ref{spectrum}c, and also \ref{phsp_part_bias} for an illustration of phase space transitions), or in the milder form of an eigenvalue ``avoided crossing,'' (inducing a smooth, yet very steep depreciation of the learned direction at a specific error, see Figure \ref{spectrum}d,g). As discussed in our previous work, these two latter effects can be practically undistinguishable: learning works reasonably well for small enough errors; for errors past the crash value, the outcome becomes irrelevant to the input statistics, and the system is essentially encoding information on the cross-talk pattern itself.

Finally, we found that in instances of highly unbiased inputs, learning may lead to an ambiguous outcome (double leading eigenvalue), even in the absence of cross-talk (Figures \ref{spectrum}f,h and \ref{phsp_full_unbias}). This is an occurrence we have not encountered in our previous, more restrictive, versions of the model, since it requires inputs with concomitant negative cross-correlations and loss of bias of order $>2$). Our current analysis shows that the fashion in which the cross-talk handles input ambiguity (i.e., nonisolated, neutrally-attracting equilibria) depends quite significantly on the number and (in this case also) geometry of the negative correlations within the input. Depending on these, we distinguished two cases. One in which even the smallest degree of cross-talk helps the system make an asymptotic selection for one particular direction in the eigenspace spanned by the multiple eigenvalue. The other, in which no small degree of inspecificity can perturb this ``stable ambiguity''. The level of critical cross-talk that can finally destroy the curve of neutrally-stable equilibria also pushes the system to learn an orthogonal direction, hence irrelevant to the main features of the original input statistics.

While this extension still only considers a very simple model of learning, it helps us re-iterate an important idea, which we have formulated before~\cite{radulescu2009hebbian,cox2009hebbian,elliott2012,radulescu2010}). A central problem for biological learning seems to be that the activity-dependent processes that lead to connection strength adjustments cannot be completely synapse specific~\cite{cox2009hebbian,adams2002new}. This raises the possibility that sophisticated learning, such as presumably occurs in the neocortex, is enabled as much by special machinery for enhancing specificity, as by special algorithms~\cite{adams2006neurobiological}. It seems therefore possible that a key biological factor in learning problems is not just in finding good architectures, techniques and estimative algorithms, but also in perfecting the relevant plasticity apparatus. We have suggested that learning plasticity errors are analogous to mutations, and that cortical circuitry might reduce such errors in the same manner as ``proofreading'' reduces DNA copying mistakes. This further suggests that problems of survival and reproduction are so diverse that no single algorithm can solve them all, so that no ``universal'' or ``canonical'' cortical circuit should be expected. However, if every specialized algorithm relies on extraordinarily specific synaptic weight adjustment, then finding machinery that allows such specificity would indeed be equivalent to discovering new neurobiological general principles.  We have speculated that an important part of such machinery, at least in the neocortex, might lie outside the synapse itself, in the form of complex circuitry performing a proofreading operation analogous to that procuring accuracy for polynucleotide copying~\cite{adams2002new,adams2006neurobiological,biomathics}. Let us note that such machinery would be less necessary if update inaccuracy merely degraded learning, rather than preventing it (possibility which our model does not exclude). Even so, when temporarily unfavorable input statistics lead to imperfect learning because of Hebbian inspecificity, the degraded weights might still be a useful starting point for better learning when input statistics improve.


\subsection*{Acknowledgements}

To my mom and dad, who took care of my world while I have been working, and to Alex, who allowed them to take over and rule the world.


\bibliographystyle{plain}
\bibliography{references}


\clearpage

\section*{Appendix 1}

The symmetric, positive definite matrix ${\bf C} \in {\cal{M}}_{n}(\mathbb{R})$ defines a dot product in $\mathbb{R}^{n}$ as:

  $$\langle {\bf v},{\bf w} \rangle_{\bf C}={\bf v}^{T}{\bf Cw}$$

Although both ${\bf C}$ and ${\bf E}$ are symmetric, the product ${\bf EC}$ is not symmetric in the Euclidean metric. However, in a new metric defined by the dot product $\langle \cdot,\cdot \rangle_{\bf C}$, ${\bf EC}$ is symmetric. (Indeed, for any pair of vectors ${\bf u},{\bf v} \in \mathbb{R}^n$, we have

$$\langle {\bf ECu},{\bf v} \rangle_{\bf C} = ({\bf ECu})^t{\bf Cv} = {\bf u}^t {\bf C}^t {\bf E}^t {\bf Cv} = {\bf u}^t {\bf CECv} = \langle {\bf u}, {\bf ECv}\rangle_{\bf C}$$

In consequence, ${\bf EC}$ has a basis of eigenvectors, orthogonal with respect to the dot product $\langle \cdot,\cdot \rangle_{\bf C}$.

\noindent The following theorem, describing the equilibria of the system (\ref{mothersys}), is immediate.\\

\noindent {\bf Theorem 1.} \begin{em}An equilibrium for the system is any vector ${\bf w}=(w_{1}...w_{n})^{T}$ such that ${\bf ECw}=({\bf w}^{T}{\bf Cw}){\bf w}$, i.e., an eigenvector of ${\bf EC}$ (with corresponding eigenvalue $\lambda_{\bf w}$), normalized, w.r.t. the norm $\lVert \cdot \rVert_{\bf C} = \langle \cdot,\cdot \rangle_{\bf C}$, so that $\lVert {\bf w} \rVert_{\bf C}=\lambda_{\bf w}$.

$${\bf ECw}=\lambda_{\bf w} {\bf w}, \quad \lVert {\bf w} \rVert_{\bf C}=\lambda_{\bf w}$$

\noindent If we additionally assume (generically) that ${\bf EC}$ has strictly positive maximal eigenvalue of multiplicity one, then the corresponding eigendirection is orthogonal in $\langle \cdot,\cdot \rangle_{\bf C}$ to all other eigenvectors of ${\bf EC}$.
\end{em}\\

\noindent Take then ${\bf w}$ to be an equilibrium of the system (\ref{mothersys}), i.e. an eigenvector of ${\bf EC}$, with eigenvalue $\lambda_{\bf w}=({\bf w}^{T}{\bf Cw}){\bf w} > 0$. To establish stability, we calculate the Jacobian matrix at ${\bf w}$ to be

$$Df^{\bf E}_{\bf w} = \gamma \left[ {\bf EC} - 2{\bf w}({\bf Cw})^{T} - ({\bf w}^{T}{\bf Cw}){\bf I} \right]$$

\noindent Then we get the following:

\noindent {\bf Theorem 2.} \begin{em} Suppose ${\bf EC}$ has a multiplicity one largest eigenvalue. An equilibrium ${\bf w}$ (i.e., by theorem (), an eigenvector of ${\bf EC}$ with eigenvalue $\lambda_{\bf w}$, normalized so that $\| w \|_{\bf C} = \lambda_{\bf w}$) is a local hyperbolic attractor for (\ref{mothersys}) iff it is an eigenvector corresponding to the maximal eigenvalue of ${\bf EC}$.
\end{em}\\

\proof{Fix an eigenvector ${\bf w}$ of ${\bf EC}$, with ${\bf ECw} = \lambda_{\bf w} {\bf w}$. Then:

\begin{eqnarray*}
Df^{\bf E}_{\bf w}{\bf w} &=& - 2 \gamma \lambda_{\bf w}{\bf w}
\end{eqnarray*}

\noindent Recall that the vector ${\bf w}$ can be completed to a basis ${\cal B}$ of eigenvectors, orthogonal with respect to the dot product $\langle \cdot,\cdot \rangle_{\bf C}$. Let ${\bf v} \in {\cal B}$, ${\bf v} \neq {\bf w}$, be any other arbitrary vector in this basis, so that ${\bf ECv}=\lambda_{\bf v} {\bf v}$, and $\langle {\bf w}, {\bf v} \rangle_{\bf C} = {\bf w}^t{\bf Cv} = 0$. We calculate:

\begin{eqnarray*}
Df^{\bf E}_{\bf w}{\bf v} &=& - \gamma [\lambda_{\bf w} - \lambda_{\bf v}] {\bf v}
\end{eqnarray*}

\noindent So ${\cal{B}}$ is also a basis of eigenvectors for $Df^{\bf E}_{\bf w}$. The corresponding eigenvalues are $-2 \gamma \lambda_{\bf w}$ (for the eigenvector ${\bf w}$) and $-\gamma[\lambda_{\bf w} - \lambda_{\bf v}]$ (for any other eigenvector ${\bf v} \in {\cal{B}}, \, , \, {\bf v} \not= {\bf w}$). An equivalent condition for ${\bf w}$ to be a hyperbolic attractor for the system (\ref{mothersys}) is that all the eigenvalues of $Df^{\bf E}_{\bf w}$ are $<0$. Since $\gamma, \lambda_{\bf w} > 0$, this condition is further equivalent to having $- \gamma(\lambda_{\bf w} - \lambda_{\bf v}) \rvert < 0 \, , \text{ for all } {\bf v} \in {\cal{B}} \, , \, {\bf v} \not= {\bf w}$. In conclusion, an equilibrium ${\bf w}$ is a hyperbolic attractor if and only if $\lambda_{\bf w} > \lambda_{\bf v} \, , \, \text{ for all }\, {\bf v} \not= {\bf w}$ \, (i.e. $\lambda_{\bf w}$ is the maximal eigenvalue, or in other words if ${\bf w}$ is in the direction of the principal eigenvector of ${\bf EC}$).

}

\noindent Such attractors always exist provided that the condition of Theorem 2 is met (i.e., ${\bf EC}$ has a maximal eigenvalue of multiplicity one). Then the network learns, depending on its initial state, one of the two stable equilibria, which are the two (opposite) maximal eigenvectors of the modified input distribution, normalized so that $\| {\bf w} \|_{\bf C} = \lambda_{\bf w}$. Next, we aim to show that these two attractors are the system's only hyperbolic attractors.\\

\noindent {\bf Theorem 3.} \begin{em} Suppose the the modified covariance matrix ${\bf EC}$ has a unique maximal eigenvalue $\lambda_1$. Then the two eigenvectors $\pm {\bf w_{EC}}$ corresponding to $\lambda_1$, normalized such that $\| {\bf w} \|_{\bf C} = \lambda_1$, are the only two attractors of the system. More precisely, the phase space is divided into two basins of attraction, of ${\bf w_{EC}}$ and $- {\bf w_{EC}}$ respectively, separated by the subspace $\langle {\bf w}, {\bf w_{EC}} \rangle = 0$.
\end{em}\\

\proof{ We make the change of variable ${\bf u}=\sqrt{\bf C} {\bf w}$. The system then becomes:

\begin{equation}
\dot{\bf u} = {\bf Au} - ({\bf u}^t {\bf u}) {\bf u}
\label{newsys}
\end{equation}

\noindent where ${\bf A = \sqrt{C} E \sqrt{C}}$ symmetric matrix, having the same eigenvalues as ${\bf EC}$. More precisely, ${\bf w}$ is an eigenvector of ${\bf EC}$ with eigenvalue $\mu$ iff ${\bf \sqrt{C} v}$ is an eigenvector of ${\bf A}$ with eigenvalue $\mu$; hence any two distinct eigenvectors of ${\bf A}$ are orthogonal in the regular Euclidean dot product.\\

\noindent Consider then ${\bf v}$ to be the leading eigenvector of ${\bf A}$, and let ${\bf u = u}(t)$ be a trajectory of the system (\ref{newsys}). We want to observe the evolution in time of the angle between the variable vector ${\bf u}$ and the fixed vector ${\bf v}$, measured as:

\begin{equation}
\cos {\theta} = \frac{\langle {\bf v}, {\bf u} \rangle}{\| {\bf v} \| \cdot |\ {\bf u} \|}
\nonumber
\end{equation}

\noindent We differentiate and obtain:

\begin{equation}
- \| {\bf v} \| \sin(\theta) \dot{\theta} = \frac{({\bf v}^t \dot{\bf u}) \| {\bf u}\|^2 - ({\bf v}^t{\bf u}) ({\bf u}^t {\bf u})}{\| {\bf u} \|^3}
\nonumber
\end{equation}

\noindent The numerator of this expression is

\begin{equation}
h({\bf u}) = ({\bf u}^t {\bf u})({\bf v}^t {\bf Au}) - ({\bf v}^t {\bf u})({\bf u}^t {\bf Au})
\end{equation}

\noindent We are interested in the sign of $h({\bf u})$; to make our computations simpler, we can diagonalize ${\bf A}$ in a basis of orthogonal eigenvectors  ${\bf A} = {\bf P}^t{\bf DP}$, where ${\bf D}$ is the diagonal matrix of eigenvalues and ${\bf P}$ is an orthogonal matrix whose columns are the eigenvectors. Then:

\begin{equation}
h({\bf u}) = ({\bf z}^t {\bf z}) ({\bf y}^t {\bf D}{\bf z}) - ({\bf y}^t {\bf z}) ({\bf z}^t {\bf D}{\bf z})
\nonumber
\end{equation}

\noindent where ${\bf y} = {\bf Pv}$ and ${\bf z} = {\bf Pu}$, so that ${\bf Dy} = {\bf DPv} = \lambda_1 {\bf y}$ (where $\lambda_1$ is the largest eigenvalue of ${\bf EC}$, assumed to have multiplicity one). Then:

\begin{equation}
h({\bf u}) = ({\bf y}^t {\bf z}) \sum_{j=2}^n{(\lambda_1 - \lambda_j) z_j^2}
\nonumber
\end{equation}

\noindent Hence, if ${\bf y}^t {\bf z} >0$, then $h({\bf u}) > 0$. In other words: if ${\bf v}^t {\bf u} > 0$ then $- \| {\bf v} \| \sin(\theta) \dot{\theta} > 0$, hence that $\dot{\theta} < 0$. For our original system, this means that any trajectory starting at a ${\bf w}$ with $\langle {\bf w}, {\bf w_{EC}} \rangle > 0$ converges in time towards the principal eigenvector ${\bf w_{EC}}$ of the matrix ${\bf EC}$.
}


\clearpage

\section*{Appendix 2}

We want a concise description of the modified input matrix ${\bf EC}$. To begin with, we can express the matrices ${\bf E}$ and ${\bf C}$ individually as: $\ds {\bf E} = \epsilon {\bf M} + (q-\epsilon){\bf I}$ and $\ds {\bf C} = c{\bf M} + (v-c){\bf I} + \sum{\delta_j {\bf A}_j}$, where ${\bf I}$ is the $n \times n$ identity matrix, ${\bf M}$ is the $n \times n$ matrix with uniform unit entries, and, for any $j=\overline{1,n}$, ${\bf A}_j$ is the matrix with zero entries except $A_j(j,j)=1$. Note, for future computations, that ${\bf M}^2=n{\bf M}$ and that ${\bf MA}_j$ is the matrix with the only nonzero entries being ones along the $j$-th column. Unless otherwise specified, the summations are for $j=\overline{1,n}$. The product ${\bf EC}$ will then be

\begin{eqnarray}
{\bf EC} &=& [\epsilon (v-c) + c(q-\epsilon) + \epsilon cn]{\bf M} + (q-\epsilon)(v-c){\bf I} + \epsilon \sum{\delta_j {\bf MA}_j} + (q-\epsilon) \sum{d_j{\bf A}_j} \nonumber
\end{eqnarray}

\noindent In matrix form, this translates as \\

${\bf EC} = \left \lvert \begin{array}{cccc}
X_1(\lambda)&f_2&\cdots &f_n\\
f_1&X_2(\lambda)&\cdots &f_n\\
\vdots & &\ddots &\vdots \\
f_1&f_2&\cdots &X_n(\lambda)\\
\end{array}\right \rvert$\\

\vspace{2mm}
\noindent where, $\forall j=\overline{1,n}$, we called $f_j=\epsilon(v+\delta_j-c)+c$ and $X_j(\lambda)=q(v+\delta_j-c)+c -\lambda$.


\subsection*{Fully biased case}

We first consider the covariance biases $\delta_j$'s to be distinct:  $\delta_1 > \delta_2 > \hdots > \delta_{n-1} > \delta_n = 0$. We will prove that the polynomial $\Delta$ has $n$ real roots $\xi_1 \geq \xi_2 \geq \hdots \geq \xi_n$, and we will find approximating bounds for their positions on the real line.\\

\noindent Remark first that the end behavior of $\Delta(\lambda)$ is given by: $\ds \lim_{\lambda \to -\infty}{\Delta(\lambda)}=\infty \label{minusinfinity}$ and $\ds \lim_{\lambda \to +\infty}{\Delta(\lambda)}=(-1)^n \infty \label{plusinfinity}$\\

\noindent Consider $\lambda_j=(q-\epsilon)(v+\delta_j-c)$; clearly: $\lambda_1 >\lambda_2> ... >\lambda_n$. We will use these values to partition the real line and separate the roots of $\Delta$. To begin, we calculate, for all $i,j=\overline{1,n}$:

\begin{eqnarray}
X_i(\lambda_j)&=& f_i+(q-\epsilon)(\delta_i-\delta_j)
\end{eqnarray}

\noindent In particular: $X_j(\lambda_j)=f_j, \; \forall j=\overline{1,n}$. By raw and column manipulations, it can be shown that, $\forall j=\overline{1,n}$

\begin{eqnarray}
\noindent
\Delta(\lambda_j)
&=& f_j (q-\epsilon)^{n-1}\prod_{i \neq j}{(\delta_i-\delta_j)}
\end{eqnarray}

\noindent In consequence: $\ds \text{sign}(\Delta(\lambda_j)) = \text{sign}(f_j) (-1)^{n-j}$.\\

\noindent Recall that $f_j=\epsilon(v+\delta_j-c)+c$, hence $f_1>f_2> \hdots > f_n$. To continue our discussion and establish the signs of $\Delta$ at all partition points $\lambda_j$, we need to establish the index $j$ for which the values $f_j$ switch sign.

For each $j \in \overline{1,n}$, consider the ``critical'' error values, for which $f_j(\epsilon_j^*)=0, \; \forall j \in \overline{1,n}$:

\begin{equation}
\epsilon_j^*= \frac{-c}{v+\delta_j-c}
\end{equation}

\noindent so that $0<\epsilon_1^*<\epsilon_2^*< \hdots < \epsilon_n^*$ (since $\delta_1>\delta_2> \hdots > \delta_n$). Clearly, for all $j \in \overline{1,n}$, we have $f_j >0$ iff $\epsilon > \epsilon_j^*$.\\

\noindent {\bf Remark.} A safe assumption that would allow us to study all cases that may appear is to consider $v > (n-1)\lvert c \rvert$, which guarantees $\epsilon_j^* < 1/n$, $\forall j \in \overline{1,n}$. This insures a complete discussion, since then $\epsilon \in [0,1/n]$ is allowed to reach and cross over all the critical values $\epsilon_j^*$, creating a possible swap in the order of the eigenvalues of ${\bf EC}$, as we will show later. The proof for the other cases will be omitted, since it is just a simplification of the present argument. In fact, the only crossover of true interest to us is $\epsilon=\epsilon_1^*$, where the eigenvalue swap involves the two largest eigenvalues and thus affects the position of the system's attracting equilibria, corresponding to the normalized eigenvectors of the maximal eigenvalue; the other critical values $\epsilon=\epsilon_j^*$, for $j \geq 2$, only affect the stable/unstable spaces of the saddle equilibria. In this light, the condition on the entries of the covariance matrix can be loosened to $v > (n-1) \lvert c \rvert -\delta_1$.\\

\noindent We distinguish the following cases:

\begin{description}
\item ({\bf I.}) For $0 \leq \epsilon < \epsilon_1^*$. This implies $f_j<0, \: \forall j \in \overline{1,n}$. Then
\begin{equation}
\text{sign}(\Delta(\lambda_j)) = \text{sign}(f_j) (-1)^{n-j} = (-1)(-1)^{n-j} = (-1)^{n-j+1}
\label{signature1}
\end{equation}

\noindent From (\ref{minusinfinity}),(\ref{plusinfinity}) and (\ref{signature1}), we obtain the following sign table:

\begin{center}
\begin{doublespacing}
\begin{tabular}{c|llllllll|}
$\lambda$ & $-\infty$ & $\lambda_n$ & $\lambda_{n-1}$ & $\hdots$ & $\lambda_2$ & $\lambda_1$ & $+\infty$ \\
\hline
$\text{sign}(\Delta(\lambda))$ & $(+)$ & $(-)$ & $(+)$ & $\hdots$ & $(-1)^{n-1}$ & $(-1)^n$ & $(-1)^n$ \\
\end{tabular}
\end{doublespacing}
\end{center}

\noindent From the Intermediate Value Theorem and the Fundamental Theorem of Algebra, it follows that the polynomial $\Delta(\lambda)$ has $n$ real roots $\xi_1 > \xi_2 > \hdots > \xi_n$, such that:

\begin{equation}
-\infty < \xi_n < \lambda_n < \xi_{n-1} < \lambda_{n-1} < \hdots < \lambda_2 < \xi_1 < \lambda_1 < \infty
\label{order}
\end{equation}

\item ({\bf II.}) For $\epsilon_p^* < \epsilon < \epsilon_{p+1}^*$. Then $f_1, \hdots , f_p >0$ and $f_{p+1}, \hdots , f_n <0$. Similarly as in ({\bf I.}), we have:

\begin{center}
\begin{doublespacing}
\begin{tabular}{c|llllllllll|}
$\lambda$ & $-\infty$ & $\lambda_n$ & $\lambda_{n-1}$ & $\hdots$ & $\lambda_{p+1}$ & $\lambda_p$ & $\hdots$ & $\lambda_1$ & $+\infty$ \\
\hline
$\text{sign}(\Delta(\lambda))$ & $(+)$ & $(-)$ & $(+)$ & $\hdots$ & $(-1)^{n-p}$ & $(-1)^{n-p}$ & $\hdots$ & $(-1)^{n-1}$ & $(-1)^n$ \\
\end{tabular}
\end{doublespacing}
\end{center}

\noindent hence the polynomial $\Delta(\lambda))$ has roots $\xi_1 > \xi_2 > \hdots > \xi_n$, such that:

\begin{equation}
-\infty < \xi_n < \lambda_n < \xi_{n-1} < \lambda_{n-1} < \hdots < \xi_{p+1} < \lambda_{p+1} < \lambda_p < \xi_p < \hdots < \lambda_1 < \xi_1 < \infty
\end{equation}

\item ({\bf III.}) For $\epsilon_n^* < \epsilon < 1/n$. Then $f_1, \hdots , f_n >0$ and we have

\begin{center}
\begin{doublespacing}
\begin{tabular}{c|llllllll|}
$\lambda$ & $-\infty$ & $\lambda_n$ & $\lambda_{n-1}$ & $\hdots$ & $\lambda_2$ & $\lambda_1$ & $+\infty$ \\
\hline
$\text{sign}(\Delta(\lambda))$ & $(+)$ & $(+)$ & $(-)$ & $\hdots$ & $(-1)^{n}$ & $(-1)^{n-1}$ & $(-1)^n$ \\
\end{tabular}
\end{doublespacing}
\end{center}

\noindent and the polynomial $\Delta(\lambda))$ has roots $\xi_1 > \xi_2 > \hdots > \xi_n$, such that:

\begin{equation}
-\infty < \lambda_n < \xi_{n} < \lambda_{n-1} < \xi_{n-1} < \hdots  < \lambda_1 < \xi_1 < \infty
\end{equation}

\end{description}

\noindent In particular, we have proved the following lemma in the main text:\\

\noindent {\bf Proposition 3.1.} \emph{In the biased case $\delta_1>\delta_2> \hdots >\delta_n$, the matrix ${\bf EC}$ has $n$ real distinct eigenvalues $\xi_1>\xi_2> \hdots > \xi_n$.}

\subsection*{Losing the bias}

\noindent Suppose now that, for $j \in \overline{1,n}$, $\delta_j = \delta_{j+1}+\zeta_j$, and allow some of the $\zeta_j \to 0$; in the limit, this results in a loss of bias in the covariance matrix ${\bf C}$ ($v+\delta_j=v+\delta_j+1$ for some index $j$). In consequence, $\lambda_j - \lambda_{j+1} \to 0$.

Let's study the changes of the maximal root $\xi_1$ as $\zeta_1 \to 0$ (i.e., we eliminate the bias between the two most correlated components of the matrix ${\bf C}$. Suppose $\epsilon \in \left[ 0,\epsilon_1^* \right]$. This calculation can be extended similarly to the other intervals for $\epsilon$; however, we will only discuss here the case $\epsilon \in \left[ 0,\epsilon_1^* \right]$, since it is the only one that relates directly to the position and multiplicity of the \emph{leading} root of $\Delta$. It also agrees with our goal to study the behavior of the system for small enough transmission errors. According to (\ref{order}), we have

$$-\infty < \xi_n < \lambda_n < \xi_{n-1} < \lambda_{n-1} < \hdots < \lambda_2 < \xi_1 < \lambda_1 < \infty$$

Since $\lambda_1 \to \lambda_2$, it follows that in the limit of $\zeta=0$ and $\xi=\lambda_1=\lambda_2$, so that the maximal eigenvalue of ${\bf EC}$ preserves its multiplicity =1. This situation changes if we introduce an \emph{order two bias loss} $\delta_1=\delta_2=\delta_3$ (i.e. if we make both $\zeta_1$ and $\zeta_2$ approach zero simultaneously). Then $\lambda_1-\lambda_2 \to 0$ and $\lambda_2-\lambda_3 \to 0$, so that, the two leading roots collide into a double root $\lambda_3=\xi_2=\lambda_2=\xi_1=\lambda_1$. This justifies the following proposition:\\

\noindent {\bf Proposition 3.2.} \emph{Suppose $\epsilon<\epsilon_1^*$. An \emph{order $k$ bias loss} of the covariance matrix ${\bf C}$ of the type $\delta_1=\hdots=\delta_k$ results in a leading eigenvalue of multiplicity $k-1$ for the modified covariance matrix ${\bf EC}$.}\\

This proposition can be generalized to encompass bias loss anywhere in the inputs, and any interval for the error $\epsilon$. Below, we give a more general statement, which follows by repeating the argument for the case we already analyzed, but could also be proved more directly.\\

\noindent {\bf Theorem.} \begin{emph} Suppose that the matrix ${\bf C}$ is allowed to exhibit bias loss in all possible ways, so that it can be written in block form as (\ref{covariance}), where there exist $k_1,k_2, \hdots ,k_N \in \overline{1,n}$, with $\displaystyle{\sum_{j=1}^{N}{k_j}=n}$ and such that

\begin{eqnarray}
\delta_1&=&\hdots=\delta_{k_1}=\nu_1 \nonumber\\
\delta_{k_1+1}&=&\hdots=\delta_{k_2}=\nu_2 \nonumber\\
&&\vdots \nonumber\\
\delta_{k_{N-1}+1}&=&\hdots=\delta_{k_N}=\nu_N \nonumber
\end{eqnarray}

\noindent with

$$\nu_1>\nu_2>\hdots>\nu_N$$

\noindent Then the characteristic polynomial $\Delta$ of ${\bf EC}$ has all real eigenvalues. More precisely, these eigenvalues are $\lambda_j=(q-\epsilon)(v+\delta_j-c)$ with multiplicity $k_j-1$, for all $j \in \overline{1,N}$, and $N$ additional eigenvalues $\xi_1 \geq \xi_2 \geq \hdots \geq \xi_N$.
\end{emph}

\vspace{2mm}
\noindent {\bf Remark.} The order of these eigenvalues, depending on the the error value $\epsilon$ with respect to the critical error values $\displaystyle{\nu_j^*=\frac{-c}{v+\nu_j-c}}$, is the same as described in the cases \emph{({\bf I.})-({\bf III.})} above.

\end{document}